\documentclass[twocolumn]{revtex4}
\usepackage{graphicx}
\usepackage[]{epsfig}

\newcommand{\beq}{\begin{equation}}
\newcommand{\eeq}{\end{equation}}
\newcommand{\beqd}{\begin{displaymath}}
\newcommand{\eeqd}{\end{displaymath}}
\newcommand{\beqa}{\begin{eqnarray}}
\newcommand{\eeqa}{\end{eqnarray}}

\newcommand{\comment}[1]{}
\newcommand{\bk}{\mathbf{k}}
\newcommand{\bq}{\mathbf{q}}
\newcommand{\bp}{\mathbf{p}}

\begin{document}

\title{Super-Cooled Liquids: Equivalence between  Mode-Coupling Theory and Replica Approach}

\author{Tommaso Rizzo$^{1,2}$}
\affiliation{$^1$ Dip. Fisica,
Universit\`a "Sapienza", Piazzale A. Moro 2, I-00185, Rome, Italy \\
$^2$ IPCF-CNR, UOS Rome, Universit\`a "Sapienza", PIazzale A. Moro 2,
I-00185, Rome, Italy}

\begin{abstract}
We show that the replica approach to glassy dynamics provides, in spite of its static nature, a characterization of critical dynamics in the $\beta$-regime of super-cooled liquids that is equivalent to the one of Mode-Coupling-Theory, both qualitatively and quantitatively. The nature and extent of this equivalence is discussed in connection to the main open problems of the current theory.
\end{abstract}
\maketitle

\section{Introduction}

It is nowadays well established that the Mode-Coupling-Theory (MCT) yields and accurate description of the early stages of  the dynamical slowing down in super-cooled glass-forming liquids \cite{Gotze09}.
One of its main successes is the prediction that the slow relaxation has a two-step nature with time correlators developing plateaus that are expected to diverge at some critical temperature characterized by a transition to a glass phase.

The main drawback of the theory is that the critical temperature is actually significantly larger than the one of the  glass-transition in numerical simulations and it seems rather to mark a dynamical cross-over \cite{Hansen}.
Nevertheless various quantities, like the so-called non-ergodicity parameter (essentially the height of the plateau of the correlators), compare very well with the numerical data. Furthermore the theory provides a detailed set of predictions for the behavior of the critical correlators and the dynamical exponents \cite{Gotze85} that reproduce quite well the numerical data \cite{Nauroth97,Sciortino01,Weysser10}.
For these reasons many believe that the inclusion of some sort of corrections to the theory, possibly taking into account finite-dimensional effects leading to some some sort of hopping processes, could extend MCT down to the laboratory glass transition. Unfortunately due to the uncontrolled nature of Sjogren's approximation \cite{Sjogren80} which is at the heart of standard MCT \cite{Bengtzelius84} it is not clear how to compute systematically corrections to it.

On the other hand it has been discovered by Kirkpatrick, Thirumalai and Wolynes \cite{Kirkpatrick87c} that a certain class of mean-fied spin-glass models exhibits the same dynamical features of MCT. This result is important for various reasons. It shows that MCT works for systems that are completely different at the microscopic level from a liquid  hinting that it might have some universal features. Furthermore it allows to include, at least phenomenologically, finite-size effects in the theory. Indeed critical dynamics in these models can be tracked back to the splitting of the equilibrium state into an exponential number of metastable states and the use of nucleation arguments on this mean-field picture has led to the formulation of the so-called Random-First-Order-Theory of the glass transition \cite{Kirkpatrick87c,Biroli04}. Another important consequence is that some features of critical dynamics of these models, namely the critical temperature and the non-ergodicity parameter, can be computed without solving explicitly the dynamics by means of an essentially static approach: the replica method. This observation has led to the application of the replica method directly in the context of liquid theory, leading to a novel set of quantitative predictions for the non-ergodicity parameter and for the critical temperature \cite{Mezard99}.

The starting point of MCT is a set of exact equations for the dynamics of the system under study. These dynamical equations are closed by means of an appropriate approximation scheme and it turns out that the solution displays a transition to a glassy phase at some critical temperature. The replica approach instead deals with purely static equilibrium quantities for the given model. Actually one considers a system of $m$ replicas of the original system and wants to compute the replicated order parameter $q_{ab}$ that describes the appropriate microscopic two-point correlation ({\it i.e.} density-density in liquids or spin-spin in spin-glasses) between two identical replicas $a,b=1\dots m$ of the system. 
Using standard statistical mechanics procedures one derives a (usually approximate)  closed equation for $q_{ab}$.  Much as in MCT the dynamical glass transition is characterized by the abrupt appearance of a novel solution of the equation of state characterized by a value of $q_{ab}$ different from the value corresponding to the high temperature phase (say the liquid or paramagnetic phase depending on the system considered). The solution depends on the number $m$ of replicas considered and for technical reason one should take the limit $m\rightarrow 1$ in order to obtain the correct result. In the structural glass context it is customary to use methods based essentially on the Hyper-netted-chain (HNC) approximation \cite{Mezard99,franz12}. These methods yield predictions for the ergodicity breaking parameter qualitatively similar but quantitatively different from those of MCT. Because of this discrepancy it has been assumed in the past that the two methods are intrinsically different. In \cite{Szamel10} Szamel has shown instead that starting from replicated Ornstein-Zernicke equations and closing them with a suitable approximation scheme one can recover within the replica method the very same quantitative results of MCT.  Therefore it can be argued that the two approaches predict the same physics and that discrepancies arise only when  non-equivalent approximation schemes are used.

Obviously the two methods cannot be completely equivalent. In principle MCT can describes dynamics on both small and large times scale while the replica method is intrinsically static. 
Therefore it is reasonable that the latter can describe a static (or rather quasi-static) quantity (like the ergodicity-breaking parameter) while it cannot obviously tell anything on short-time dynamics.
Previously it was assumed that it could not tell anything on dynamics at all. Instead it has been recently realized \cite{noi}  that it can be used to characterize the critical behavior on the large time scale of the $\beta$ regime, {\it i.e.} the plateau in the time-correlators.
In the context of MCT it was realized long ago \cite{Gotze85} that this regime has essentially a universal nature in the sense that its properties are qualitatively model-independent. Besides they depend quantitatively on the given model only through a single quantity, the so-called parameter exponent $\lambda$. The very same universal equations has been derived within the replica method in \cite{noi}. In both derivations universality follows essentially from an argument {\it a la} Landau and therefore it is quite robust. 
Besides these universal results, in \cite{noi} two novel results were derived concerning the non-universal parameter exponent $\lambda$. As will be explained in the next section the first one provides a recipe to compute $\lambda$ within the replica approach, extending the amount of information that can be extracted from existing and future replica computations. The second one is more deep and states that the $\lambda$ is equal to the ratio of two quasi-static six-point susceptibilities that can in principle be observed in experiments. 

The above recipe has been applied to a number of mean-field spin-glass models yielding new analytical predictions concerning critical dynamics \cite{calta1,calta2,ferra1,calta3}. When there was already an explicit solution of the dynamics the novel computations offered an {\it a posteriori} validation of the recipe, while in other cases the new predictions were compared with existing numerical simulations. More recently the recipe has been used in the context of the replicated theory of super-cooled liquids based on the HNC approximation \cite{franz12} and for a wide range of models a (quite reasonable) value of $\lambda \approx .7$ was obtained.

In the context of MCT an equivalent recipe exists in order to compute the parameter exponent from the value of a more general object, {\i.e.} the mode-coupling functional ${\tilde F}[f(k)]$ \cite{Gotze85}.  In both MCT and the replica approach it can be argued that the recipes are actually {\it exact}, therefore in order to show that the two methods are consistent with each other we have to show that quantitative differences in the actual value of $\lambda$ arise only because different approximation schemes are used in the estimate of the mode-coupling functional in MCT or correspondingly the equation of state for $q_{ab}$ in the replica approach. Correspondingly if equivalent approximation schemes are used within the two approaches then the same (approximate) expression for $\lambda$ must be obtained. The main result presented in this paper is that once the replica recipe is applied to Szamel's closure the very same quantitative predictions of MCT are obtained, proving that MCT and the replica method are equivalent both qualitatively {\it and} quantitatively.

The issue of the equivalence between the two approaches is not an academic one.
On one hand MCT predictions are more accurate at present than those obtained with the HNC approximation, at least in three dimension. On the other hand there are situations in which the replica method may have considerable advantages. Indeed in order to derive many quantities within MCT one takes the large-time limit in the dynamical equations. The replica approach can be viewed as a compact way to describe the algebra of this limit where different replicas correspond to configurations visited dynamically at distant times. Certainly the main open problem concerning MCT is the systematic inclusion of corrections and the role finite-dimensional effects/finite-size effects \cite{Franz11b,Berthier12}. The replica approach offers the possibility to study this phenomena in a field-theoretic framework that is definitively simpler than the dynamical one. Surprisingly it has been recently discovered that the loop corrections to this theory can be controlled at all orders and a series of novel unexpected predictions have been obtained \cite{Franz11b}. However this study confirmed that the phase transition predicted by MCT is only a dynamical cross-over and therefore the purely static replica approach, as expected, is inconsistent beyond perturbation theory. The theory needs therefore to be supplemented with some form of critical dynamics but unfortunately in spite of important progresses \cite{Berthier07} we are still lacking full control of the critical behavior of the dynamical propagators, like the well-known $\chi_4$ susceptibility.

The paper is organized as follows. In section \ref{betagen} we will recall the predictions of MCT and the replica approach concerning the $\beta$-regime and illustrate the recipe obtained in \cite{noi}. In section \ref{comp} we will present our computation showing that the two methods leads to the same results.
In section \ref{conclusions} we will give our conclusions.

\section{Theory of the $\beta$-regime within MCT and the Replica Approach}
\label{betagen}

The central quantity of MCT is the normalized autocorrelation function of density fluctuations at given wave-vector ${\mathbf k}$
\beq
\Phi(k,t)\equiv \langle \delta\rho^*({\mathbf k},t)\delta\rho({\mathbf k},0)\rangle/S(k)
\eeq
where $S(k)\equiv\langle |\delta\rho(k,0)|^2\rangle$ is the static structure factor.
Within MCT dynamical equations for $\Phi(\bk,t)$ are obtained. The key feature of these equations is that below the critical temperature $T_{c}$ they predict that the long-time limit of the correlator is no longer zero (corresponding to the liquid phase) but becomes positive, $\lim{t \rightarrow \infty} \, \Phi(\bk,t)=f(k) \neq 0$, meaning that the system is in a glassy phase.

For temperatures near the critical temperature one identifies the $\beta$-regime corresponding to time-scales over which the correlator is almost equal to $f(q)$. In the liquid phase ($T>T_c$) this regime is followed by the $\alpha$-regime during which the correlator decay from $f(k)$ to zero. In the $\beta$-regime the time-dependence of the correlator is controlled by the following scaling law \footnote{In the following presentation we will follow closely G{\"o}tze's original paper \cite{Gotze85}.}:
\beq
\Phi(k,t)=f(k)+|\tau|^{1/2} f_{\pm}(t/\tau_\beta) \, \xi_c^R(k)
\label{scalfor}
\eeq
where $\tau$ is a linear function of $T_c-T$, {\it i.e.} it is negative in the liquid phase and positive in the glassy phase, correspondingly the scaling functions $f_{+}(x)$ is to be used in the glassy phase while $f_-(x)$ has to be used in the liquid phase.
The function $f_{\pm}(x)$ obeys the scale-invariant equation:
\beq
\pm 1=f_{\pm}^2(x)\left(1-\lambda\right) +\int_0^x (f_{\pm}(x-y)-f_{\pm}(x))\dot{f}_{\pm}(y)dy
\label{SVDYN2}
\eeq
For small values of $x$ both the functions $f_{\pm}(x)$ diverge as $1/x^a$, while for large values of $x$ $f_+(x)$ goes to a constant while $f_-(x)$ diverges as $-x^b$ where the exponents $a$ and $b$ are determined by the so-called parameter exponent $\lambda$ according to:
\beq
\lambda={\Gamma^2(1-a)\over\Gamma(1-2a)}={\Gamma^2(1+b)\over\Gamma(1+2b)}
\label{lambdaMCT}
\eeq
The parameter exponent $\lambda$ controls also the time scale of the $\beta$ regime that diverges with $\tau$ from both sides as $\tau_\beta \propto |\tau|^{-1/(2\,a)}$ with an unknown model-dependent factor. 

The above expressions display a great deal of universality.
In order to be quantitative one needs to specify the so-called Mode-Coupling Functional $\tilde{F}_k[\{f(q)\}]$ which depends on the model considered. The knowledge of the Mode-Coupling functional allows to determine quantitatively the critical temperature $T_c$, the ergodicity-breaking parameter  $f(q)$, the corrections $\xi_c^R(k)$ and the parameter exponent $\lambda$.
The ergodicity breaking parameter is determined indeed by the following equation:
\beq
{ f(k) \over 1-f(k)}=\tilde{F}_k[\{f(q)\}]
\label{fq}
\eeq
The other quantities depend on the first and second derivative of the Mode-Coupling functional evaluated at $f(k)$.
From now on we specialize to the two-mode case which corresponds to the assumption that the Mode-Coupling functional is quadratic in its arguments:
\beq
\tilde{F}_k[\{f(q)\}]={1 \over 2}\sum_{\bk_1,\bk_2} V^{(2)}(\bk;\bk_1,\bk_2)f(k_1)f(k_2)
\label{quad}
\eeq 
We define then:
\beq
C_{\bk,\bq} \equiv \sum_{p}V^{(2)}(\bk;\bq,\bp)f(p)(1-f(q))^2 
\eeq
With the above definitions the critical temperature is determined by the condition that eq. (\ref{fq}) admits a non-zero solution {\it and}
the linear operator $C_{\bk,\bq}$ has a (critical) eigenvalue equal to one, whose corresponding right eigenvector is precisely  the quantity $\xi_c^R(k)$ appearing in (\ref{scalfor}):
\beq
\xi_c^R(\bk)=\sum_{\bq} \, C_{\bk,\bq}\xi_c^R(\bq)
\label{tcMCT}
\eeq
Normalizing the left and right eigenvector of $C_{\bk,\bq}$ as:
\beq
\int d\bk \, \xi_c^R(\bk) \, \xi_c^L(\bk)=1\, ,
\eeq
\beq
\int d\bk \, \xi_c^R(\bk) \xi_c^R(\bk) \xi_c^L(\bk) (1-f(\bk))=1\, ,
\eeq
the parameter exponent instead turns out to be related to the second derivative of the Mode-Coupling functional, its expression being:
\beqa
\lambda & = & {1 \over 2} \int d \bk d\bq d\bp \, \xi_c^L(k) V^{(2)}(\bk,\bq,\bp)\times
\nonumber
\\
& \times & (1-f(q))^2(1-f(p))^2\xi_c^R(q)\xi_c^R(p)\ .    
\label{finlagotze}
\eeqa
As we already said eqs.  (\ref{scalfor}), (\ref{SVDYN2}) and (\ref{lambdaMCT}) are considered exact while in order to compute $f(q)$, $T_c$ and $\lambda$ one nees to specify the Mode-Coupling-Functional.
It is well-known that very good results are obtaining considering a quadratic Functional as in (\ref{quad}) with $V^{(2)}(\bk;\bq,\bp)$ given by Sjogren expression \cite{Sjogren80}: 
\beq
V^{(2)}(\bk;\bq,\bp)=
{n S(k)S(q)S(p) \over 2 k^2 \, ( \pi)^3} \delta[\bk-\bp-\bq](\hat{\bk} \cdot [\bq c(q)+\bp c(p)])^2
\label{MCTvertex}
\eeq
where $n$ is the density.

By contrast with MCT the replica approach is purely static. 
In MCT one is interested in the ergodicity breaking parameter defined as the  long-time limit of the correlator, in this limit one can assume that the configuration at time $t$ has lost correlation with that at time $t=0$. In the replica approach this limit is somehow taken from the beginning by considering a two-point function not at distant times but rather between two real uncorrelated replicas of the system. Therefore time is removed from the problem and one is left with a purely statical problem.
The central object of the theory becomes thus a replicated object. In the original spin-glass literature the order parameter is typically a global quantity, {\it e.g.} the overlap between two configurations:
\beq
q_{ab}=\sum_{i=1}^N \overline{\langle s_i^a s_i^b \rangle}
\label{replicacorr}
\eeq
where $s_i^a$ are spin at site $i$ of the replica $a$, the angle brackets mean thermal averages and the overline mean disorder average.
In the super-cooled liquid theory the natural analog of the spin overlap would be the integral of the density-density fluctuations over space:
\beq
\Phi_{ab} \equiv {\int d \bk \langle \delta\rho_a^*({\mathbf k})\delta\rho_b({\mathbf k})\rangle \over \int d \bk S(k) }
\eeq
The replicated two-point correlator $q_{ab}$ (or equivalently $\Phi_{ab}$ in supercooled liquids) obeys his own equation of state that can be computed approximately by means of many different statistical physics methods.
In the high temperature phase the system is in the paramagnetic (liquid) phase characterized by a vanishing $q_{ab}$. Much as in MCT the transition temperature is characterized by the abrupt appearance of a solution of the equation of state with a non-zero $q_{ab}=q$. In \cite{noi} it has been shown that the replica method allows to compute not only $q$ and the critical temperature but also the parameter exponent $\lambda$.
In order to do so one has to  perform an expansion of the equation for the order parameter $q_{ab}$ around the solution $q_{ab}=q$  corresponding to the glassy phase.  
Independently of the approximate method used in order to obtain the equation it follows the  we should get an equation in powers of $q_{ab}=q+\delta q_{ab}$ of this form:
\beqa
0 & = & \tau + m_2 \left(\sum_c \delta q_{bc}+\delta q_{ac}\right)+m_3
\sum_{cd}\delta q _{cd}+
\nonumber
\\
& + &  w_1 (\delta q^2)_{ab} 
+ w_2 \delta q_{ab}^2+O(\delta q^3)\ .
\label{disa}
\eeqa
where $\tau$ is linear in $T_c-T$ corresponding to the fact that $\delta q_{ab}=0$ at $\tau=0$.
In general in the r.h.s of the above equation we should have also a term of the form $m_1\, \delta q_{ab}$ which is missing because the critical temperature is determined precisely by the condition $m_1=0$ in order that the solution of the above equation ceases to exist for negative $\tau$ (corresponding to the liquid phase).  
The other coefficients $m_2$, $m_3$, $w_1$ and $w_2$ are instead finite and depend on the given model and on the external parameters.
It can be shown that the dynamics at long time can be mapped to the replicated equation of state and that the above equation (\ref{disa}) leads precisely to the scaling form of MCT described in eq. (\ref{scalfor}). 
More precisely, the dynamical analog of the replicated order parameter (\ref{replicacorr}) defined as
\beq
C(t) \equiv \sum_{i=1}^N \overline{\langle s_i(0) s_i(t) \rangle} \, ,
\eeq
exhibits a $\beta$-regime given by:
\beq
C(t)=q+|\tau|^{1/2} f_{\pm}(t/\tau_\beta)
\eeq
where the functions $f_{\pm}(x)$ and the time-scale $\tau_\beta$ obey the very same equations (\ref{SVDYN2}) with the parameter exponent given by the ratio between the quadratic coupling constant appearing in the equation of state eq. (\ref{disa}):
\beq
\lambda={w_2 \over w_1}\ .
\label{lambdanoi}
\eeq
The above result provides therefore a simple recipe to determine $\lambda$ within the replica approach: one has to obtain a closed equation for the non-ergodicity parameter using his preferred approximation scheme, expand the equation around the critical value of $q_{ab}$ up to second order as in eq. (\ref{disa}), read the coefficients $w_2,w_1$ and finally identify $\lambda$ with their ratio.

\section{Computation of the parameter exponent}
\label{comp}

In this section we will compute the parameter exponent within the replica approach as formulated by Szamel in  \cite{Szamel10}. Our starting point is the replicated Ornstein-Zernicke equations: 
\beq
\delta n_{ab}(\mathbf{k})=n^2c_{ab}(\mathbf{k})+n\sum_c  c_{ac}(\mathbf{k})\delta n_{cb}(\mathbf{k})
\label{init}
\eeq
where $\delta n_{ab}(\mathbf{k})$ is the Fourier transform of $\delta n_{ab}(\mathbf{r_1,r_2})=n_{ab}(\mathbf{r_1,r_2})-n^2$, {\it i.e.} the non trivial part of the two-particle density, $n$ being the total particle density. 
The above equations can be closed  provided an expression for the direct correlations $c_{ab}(\mathbf{k})$ in terms of the $\delta n_{ab}(\mathbf{k})$ itself is given.
The single replica correlation functions are associated by the definition to the static structure factor according to:
\beqa
\delta n_{aa}(\mathbf{k})=n (S(k)-1)
\label{SQ}
\eeqa
Under the assumption of Replica-Symmetry (RS) the equation for the diagonal components factorizes  from that of the off-diagonal components in the limit $m\rightarrow 1$ and equation (\ref{init}) reduces to 
\beq
\delta n_{aa}(\bk)=n^2 c_{aa}(\bk)S(k)
\eeq
the above equation combined with the definition (\ref{SQ}) can be seen as a definition of the direct correlation function:
\beq
c_{aa}(\bk)=c(k) \equiv {S(k)-1 \over n\, S(k) }
\eeq
Indeed the static structure factor is usually considered an input in MCT computations.
Szamel has shown that that making an appropriate set of approximations in the context of the Yvon-Born-Green hierarchy one obtains an expression for the off-diagonal direct correlation identical to Sjogren's vertex, see  eq. (\ref{MCTvertex}).
More precisely if we {\it define} the replicated non-ergodicity parameter $f_{ab}(\bk)$ according to:
\beq
\delta n_{ab}(\bk)=n \, f_{ab}(k) S(k)
\eeq
we obtain the expression:
\beq
c_{ab}(\bk)=\int d\bq d\bp \, G(\bk,\bq,\bp)f_{ab}(q)f_{ab}(p)
\eeq
where the function $G(\bk,\bq,\bp)$ has precisely Sjogren's form:
\beq
G(\bk,\bq,\bp)={1 \over 2 k^2 \, (2 \pi)^3} \delta[\bk-\bp-\bq]S(q)S(p)(\hat{\bk} \cdot [\bq c(q)+\bp c(p)])^2
\label{defG}
\eeq
Now we multiply eq. (\ref{init}) by the inverse of the matrix $n I + \delta n_{ac}(\bk)$, ($I$ is the identity in replica and momentum space), and we use the above equation to obtain:
\beq
\left(\delta n \over n \, I+ \delta n\right)_{ab}(\bk)=n \,\int d\bq d\bp \, G(\bk,\bq,\bp)f_{ab}(q)f_{ab}(p) .
\label{EQUA}
\eeq
The above equation corresponds essentially to the equation of state for the replicated order-parameter $\delta n_{ab}(\bk)$. According to the replica recipe described in the previous section we have to expand this equation around the RS solution, therefore we rewrite:
\beq
\delta n_{ab}(\bk)=\delta n_{ab}^{RS}(\bk)+n S(k)(1-f(k))^2\delta q_{ab}(\bk)
\label{npq}
\eeq
where $\delta n_{ab}^{RS}(\bk)$ has a RS structure
\beqa
\delta n_{aa}^{RS}(\bk) &= & n\,(S(k)-1)
\\
\delta n_{ab}^{RS}(\bk) &= & n\,f(k)S(k) \ \ \ a\neq b
\eeqa
The factor $n S(k)(1-f(k))^2$ has been used in the definition in order to make contact with the formulas of ref. \cite{Gotze85} reproduced in the previous section.
In order to perform the expansion of eq. (\ref{EQUA}) in powers of $\delta q_{ab}(\bk)$ we first note that its r.h.s. (where the crucial approximations concerning the MCT functional are made) has a simple diagonal structure in replica space and only the l.h.s. has a non-trivial structure.
The zero-th order term of the l.h.s. of (\ref{EQUA}) can be computed using standard properties of replica symmetric matrices and leads after some algebra to:
\beq
\left(\delta n \over n \, I+ \delta n\right)_{ab}(\bk)={f(k) \over S(k)(1-f(k))}+O(\delta q)\,, \ a \neq b
\eeq
In order to compute the next terms we rewrite it as an expansion of in powers of $\delta n_{ab}(\bk)$:
\beq
\left(\delta n \over n \, I+ \delta n\right)_{ab}(\bk)=-\sum_{j=1}^\infty (-\delta n/n)^j_{ab}(\bk)
\label{LHSEQUA}
\eeq
where the various powers  $(-\delta n/n)^j_{ab}(\bk)$ are intended in matrix sense. After the substitution the form 
(\ref{npq}) into these products and neglecting all terms where $\delta q_{ab}(\bk)$ has a power higher than two we are left with the evaluation of three possible classes of matrix products: 1) $(\delta n_{RS})^r_{ab}$, 2) $(\delta n_{RS}^r \,  \delta q \, \delta n_{RS}^s)_{ab}$ and 3) $(\delta n_{RS}^r \,\delta q \, \delta n_{RS}^s \, \delta q \, \delta n_{RS}^t)_{ab}$ for general integer $r,s,t$ (all products and powers are intended in matrix sense and we have not written explicitly the replica and Fourier indexes).
At linear order the expansion generates terms of three types: $m_1 \delta q_{ab}$, $m_2 \sum_c(\delta q_{ac}+\delta q_{bc})$ and $m_3 \sum_{cd}\delta q_{cd}$. However as we said in the previous section, we are interested in  $m_1 \delta q_{ab}$ because  the transition temperature is specified by the condition $m_1=0$ while the other terms controls quantitatively the fluctuations \cite{Franz11b} but are not relevant for the present discussion and will not be displayed in the final result.
At variance with the remaining two terms, the term $m_1 \delta q_{ab}$ depends simultaneously on both indexes $a$ and $b$. As a result its computation is simpler because since both indexes $a$ and $b$ must be explicitly present in the final result every time we have a multiplication by $\delta n_{RS}(k)=n(S(k)(1-f(k))-1)\delta_{ab}+f(k)S(k)$ we are forced to take the term that contains the Kronecker delta. The same observation is correct also at quadratic order in $\delta q(\bk)$, {\it i.e.} for the term $\sum_c \delta q_{ac}(\bk)\delta q_{cb}(\bk)$. 
These technical arguments justify the fact that, in order to compute $m_1$ and $w_1$, we can replace (\ref{npq}) with the much simpler form $\delta n_{ab}(\bk)=n(S(k)(1-f(k))-1)\delta_{ab}+n S(k)(1-f(k))^2\delta q_{ab}(\bk)$ into the l.h.s. of eq. (\ref{EQUA}) greatly simplifying the computation. 
With this prescription we finally  obtain:
\beqa
0 & = &  -{f(k) \over (1-f(k))}-\delta q_{ab}(\bk)+ (1-f(k))(\delta q^2)_{ab}(\bk)+
\nonumber
\\
& + & n S(k) \int d\bq d\bp \, G(\bk,\bq,\bp)f(q)f(p)+
\nonumber
\\
& + &  2 n S(k) \int d\bq d\bp \, G(\bk,\bq,\bp)(1-f(q))^2\delta q_{ab}(\bq)f(p)+
\nonumber
\\
& + &  n S(k) \int d\bq d\bp \, G(\bk,\bq,\bp)\times
\nonumber
\\
& \times & (1-f(q))^2(1-f(p))^2\delta q_{ab}(\bq)\delta q_{ab}(\bp) 
\label{EQUA2}
\eeqa
The equation at zero-th order determines the non-ergodicity parameter while the next orders determine the critical temperature and the parameter exponent. 
This is the same equation obtained by Szamel and it is quantitatively identical to the one found in MCT, as can be seen using eqs. (\ref{defG}), (\ref{fq}), (\ref{quad}) and (\ref{MCTvertex}).
The critical temperature is fixed by the condition that there is a direction for $\delta q_{ab}(k)$ such that the coefficient $m_1$ of the linear term vanishes.
This corresponds to the condition that the following linear operator has  a zero eigenvalue $\mu=0$:
\beqa
\mu \, \xi_c^R(k) & = & - \xi_c^R(k)
+ 2 n S(k) \int d\bq d\bp \, G(\bk,\bq,\bp)\times
\nonumber
\\
& \times & (1-f(q))^2\xi_c^R(q)f(p)
\label{eigen}
\eeqa 
where $\xi_c^R(k)$ is the right eigenvector corresponding to the zero eigenvalue.
This condition is also identical to the condition eq. (\ref{tcMCT}) encountered in MCT where it is obtained requiring that the derivative with respect to $f(q)$ of eq. (\ref{fq}) vanishes.
Here instead it follows from the condition $m_1=0$,  technically this equivalence between two different conditions is a consequence of the $m \rightarrow 1$ replica limit and of the fact that in this limit the so-called replicon and longitudinal eigenvalue are degenerate leading to many non trivial features concerning fluctuations \cite{Franz11b}.
The behavior at the critical point is controlled by the critical eigenvector $\xi_c^R(k)$ of (\ref{eigen}) and we may write at leading order:
\beq
\delta q_{ab}(k)=\delta q_{ab} \, \xi_c^R(k) \ .
\eeq
Note the similarity of the above expression with (\ref{scalfor}) that makes apparent that different replicas play essentially the role of different times.
The above definition must be supplemented with a normalization condition for the eigenvector, in order to make contact with the result of the previous section we adopt the following normalization for the right and left eigenvectors:
\begin{displaymath}
\int d \bk  \, \xi_c^L(k) \xi_c^R(k)=1\, ,  \int d \bk  \, \xi_c^L(k) (1-f(k))\xi_c^R(k)\xi_c^R(k)=1\, .
\end{displaymath}
Multiplying the quadratic part of equation (\ref{EQUA2}) by the left eigenvector $\xi_c^L(k)$ and integrating we obtain the following expression:
\begin{widetext}
\beq
\left[ \int d \bk  \, \xi_c^L(k) (1-f(k))\xi_c^R(k)\xi_c^R(k)\right] (\delta q_{ab})^2+n \left[ \int d \bk d\bq d\bp \, \xi_c^L(k) S(k)G(\bk,\bq,\bp)(1-f(q))^2(1-f(p))^2\xi_c^R(q)\xi_c^R(p)\right] \delta q_{ab}^2\ ,
\eeq
\end{widetext}
Note that the first coefficient is fixed to one by the normalization. According to eq. (\ref{lambdanoi}) the parameter exponent is given by the ratio between the two quadratic coefficients leading to:
\beqa
\lambda & = & n \int d \bk d\bq d\bp \, \xi_c^L(k) S(k)G(\bk,\bq,\bp)\times
\nonumber
\\
& \times & (1-f(q))^2(1-f(p))^2\xi_0^R(q)\xi_0^R(p)    
\label{finla}
\eeqa
Using eq. (\ref{defG}) we see that the above expression is identical to (\ref{finlagotze}) with the vertex given by (\ref{MCTvertex}).

\section{Conclusions}
\label{conclusions} 
The recipe of  \cite{noi} has been applied to the approximation scheme proposed by Szamel in order to show that  the replica approach gives a characterization of the $\beta$-regime of super-cooled liquids that is equivalent to the one of MCT both qualitatively {\it and} quantitatively, provided the same approximations are used. 
Quantitative differences within the predictions of the two methods, see {\it e.g.} the results of \cite{franz12}, may arise only as a consequence of using approximations schemes that are not equivalent.

We note that a novel result obtained within the replica approach is that an equation like (\ref{disa})  follows from a replicated free energy \cite{noi}. This allows to connect the coefficients $w_1$ and $w_2$ with two six-point susceptibilities $\omega_1$ and $\omega_2$ which are in principle measurable in a quasi-static framework. According to \cite{noi} their ratio is equal to $\lambda$ and therefore eq. (\ref{finla}) fixes their value up to an unknown constant. 
The computation of this constant is left for future work. In the present framework it could be determined considering the response of the non-ergodicity parameter to a random pinning field along the lines of \cite{Biroli06}.  

{\em Acknowledgments.} ~~ I thank G. Szamel for useful discussions. The European Research Council has provided financial support through ERC grant agreement no. 247328.

\end{document}